# Tailoring a two-dimensional electron gas at the LaAlO$_3$/SrTiO$_3$ (001) interface by epitaxial strain

C. W. Bark[a], D. A. Felker[b], Y. Wang[c], Y, Zhang[d], H. W. Jang[a], C. M. Folkman[a], J. W. Park[a], S. H. Baek[a], X. Q. Pan[d], E. Y. Tsymbal[c], M. S. Rzchowski[b], C. B. Eom[a,1]

[a]Department of Materials Science and Engineering, University of Wisconsin, Madison, WI 53706, USA

[b]Department of Physics, University of Wisconsin, Madison, WI 53706, USA

[c]Department of Physics and Astronomy, Nebraska Center for Materials and Nanoscience, University of Nebraska, Lincoln, NE 68588, USA

[d]Department of Materials Science and Engineering, University of Michigan, Ann Arbor, MI 48109, USA

[1]To whom correspondence should be addressed. E-mail: eom@engr.wisc.edu.

**Corresponding author**: Chang-Beom Eom,
2164 ECB, 1550 Engineering Drive, Madison, WI 53706
Tel: 608/263-6305, Fax: 608/263-9017, E-mail: eom@engr.wisc.edu

**Abstract**

Recently a metallic state was discovered at the interface between insulating oxides, most notably $LaAlO_3$ and $SrTiO_3$. Properties of this two-dimensional electron gas (2DEG) have attracted significant interest due to its potential applications in nanoelectronics. Control over this carrier density and mobility of the 2DEG is essential for applications of these novel systems, and may be achieved by epitaxial strain. However, despite the rich nature of strain effects on oxide materials properties, such as ferroelectricity, magnetism, and superconductivity, the relationship between the strain and electrical properties of the 2DEG at the $LaAlO_3/SrTiO_3$ heterointerface remains largely unexplored. Here, we use different lattice constant single crystal substrates to produce $LaAlO_3/SrTiO_3$ interfaces with controlled levels of biaxial epitaxial strain. We have found that tensile strained $SrTiO_3$ destroys the conducting 2DEG, while compressively strained $SrTiO_3$ retains the 2DEG, but with a carrier concentration reduced in comparison to the unstrained $LaAlO_3/SrTiO_3$ interface. We have also found that the critical $LaAlO_3$ overlayer thickness for 2DEG formation increases with $SrTiO_3$ compressive strain. Our first-principles calculations suggest that a strain-induced electric polarization in the $SrTiO_3$ layer is responsible for this behavior. It is directed away from the interface and hence creates a negative polarization charge opposing that of the polar $LaAlO_3$ layer. This both increases the critical thickness of the $LaAlO_3$ layer, and reduces carrier concentration above the critical thickness, in agreement with our experimental results. Our findings suggest that epitaxial strain can be used to tailor 2DEGs properties of the $LaAlO_3/SrTiO_3$ heterointerface.

\body

**Introduction**

Strains have previously been used to engineer and enhance numerous properties of materials. For example, increased mobility in semiconductors (1,2), and increased transition temperature in ferroelectric materials (3,4,5,6) and superconductors (7) have been observed. A recently discovered two-dimensional electron gas (2DEG) at the $LaAlO_3/SrTiO_3$ interface (8,9) has attracted great interest due to its novel application to nanoscale oxide devices (10). So far, most studies of 2DEGs at oxide interfaces were performed using $TiO_2$-terminated $SrTiO_3$ bulk single crystal substrates. Thus, despite the rich nature of strain effects on oxide materials properties, the relationship between the strain and electrical properties of the 2DEG at the $LaAlO_3/SrTiO_3$ heterointerface remains largely unexplored.

The importance of strain effects comes from the fact that integrating 2DEGs to other functional devices or substrates always involves strain. Therefore, it is desirable to know the effect of the strain on 2DEG at the LaAlO$_3$/SrTiO$_3$ interface. In addition, by changing strain we might be able to obtain novel functional properties. For example, strain can induce an electric polarization in otherwise non-polar SrTiO$_3$ (11). It has been predicted that polarization can be used to control 2DEG properties at oxide heterointerfaces (12,13)  Thus by using the relation between the polarization and the strain, we could engineer 2DEG behavior.

To address these outstanding issues, we explore the effect of epitaxial strain on conductive properties of LaAlO$_3$/SrTiO$_3$ interface. We create the 2DEG interface on strained single-crystal (001) SrTiO$_3$ templates grown on perovskite oxide substrates with various lattice mismtach. Pseudomorphic growth of the LaAlO$_3$/SrTiO$_3$ bilayer produces a continuously strained system, including the interface at which the 2DEG resides. This allows us to add a new degree of freedom in the LaAlO$_3$/SrTiO$_3$ system and investigate the strain effect on its transport properties. We demonstrate that tensile strain makes the interface insulating, while compressive strain makes the interface metallic and allows modulating the critical thickness of LaAlO$_3$ and conductivity of the 2DEG.

**Sample growth**

LaAlO$_3$/SrTiO$_3$ thin films heterostructrues were grown on various single crystal substrates using pulsed-laser deposition (PLD) with *in-situ* high pressure reflection high-energy electron diffraction (RHEED) (14).   Figure 1A shows the schematic of the thin film heterostructure. Table 1 shows substrates that were used in this study to vary the SrTiO$_3$ strain state from biaxial compressive to biaxial tensile in the plane. As shown in figure 1, (001) SrTiO$_3$ thin films were grown on (110) NdGaO$_3$ (NGO), (001) (LaAlO$_3$)$_{0.3}$–(Sr$_2$AlTaO$_3$)$_{0.7}$ LSAT, (110) DyScO$_3$ (DSO) and (110) GdScO$_3$ (GSO) substrates. The varying lattice parameters result in an average biaxial strain ranging from -1.21% (compressive) to +1.59% (tensile) in a fully commensurate SrTiO$_3$ deposited film. All grown single-crystal (001) SrTiO$_3$ templates were fully coherent with the substrates. (001) SrTiO$_3$ films were also grown on (001) silicon substrates using Molecular Beam Epitaxy. Thickness of these quasi-single-crystal (001) SrTiO$_3$ templates on silicon was 100 nm, and the films were almost fully relaxed. The measured SrTiO$_3$ lattice parameters on Si correspond to an average biaxial strain of 0.15% (15,16). The bi-axial strain state and lattice parameters of the strained (001) SrTiO$_3$ templates are summarized in Table 1. The full width at half maximum (FWHM) values of 002 rocking curves for the strained SrTiO$_3$ template are much narrower than that of the bulk SrTiO$_3$ single crystal (17). The single-crystal (001) SrTiO$_3$ templates were also etched using buffered HF solution to maintain Ti-termination after the growth. The atomic percent of Sr, Ti, O in the films were determined with wavelength dispersive x-ray spectroscopy (WDS). The chemical ratio of grown templates was the same as the one of SrTiO$_3$ bulk single crystal within an error range. This confirms that the quality of SrTiO$_3$ templates is comparable with the bulk single crystal SrTiO$_3$ substrate, therefore, we rule out any extrinsic effects in our experiments.

LaAlO$_3$ overlayers were deposited using PLD on these various strain Ti-terminated single crystal SrTiO$_3$ templates. RHEED intensity oscillations of the specular spots show layer-by-layer growth mode as observed for LaAlO$_3$ films on SrTiO$_3$ single crystal substrates as shown in figure 1B. High resolution TEM cross-sectional analysis in figure 1C shows the LaAlO$_3$/SrTiO$_3$ film on LSAT with high crystalline quality and atomically sharp interface. In all strain states, surfaces of LaAlO$_3$ and SrTiO$_3$ films were atomically smooth with single unit cell high steps measured by atomic force microscopy (AFM), as seen in figure 1D. As a result, we confirmed that all biaxial strained heterostructures in this report were atomically controlled and grown epitaxially. More details about growth are described in ref. (11) and *Materials and Methods*.

**Results and Discussion**

It is known experimentally that a conducting 2DEG forms at the LaAlO$_3$ / bulk SrTiO$_3$ interface only after the LaAlO$_3$ overlayer thickness exceeds a critical value of 4 unit cells (18). We have found that this critical thickness depends on the strain of the system. We determined this by measuring the conductivity of strained LaAlO$_3$/SrTiO$_3$ bilayers for different thickness of the LaAlO$_3$ layer. As shown in Fig. 1, LaAlO$_3$ overlayer thickness was changed from 0 to 30 unit cells while the thickness of SrTiO$_3$ template on NGO, LSAT, DSO, GSO substrates was fixed at 50 unit cells. We also checked the critical thickness of LaAlO$_3$ on Ti-terminated (001) SrTiO$_3$ bulk single crystal and on quasi-single-crystal (001) SrTiO$_3$ templates on silicon (19) as a reference.

In case of the two samples with un-strained SrTiO$_3$ layers ( LaAlO$_3$ on bulk single crystal SrTiO$_3$ substrate and LaAlO$_3$ on relaxed SrTiO$_3$ templates on silicon), the critical thickness was in agreement with that previously reported, i.e. 4 unit cells. However, in the compressive strain states, (SrTiO$_3$ templates on LSAT and NGO), the critical thickness of LaAlO$_3$ increased to 10 unit cells and 15 unit cells, respectively, as shown in Fig. 2A. In all the cases, the conductivity saturated above the critical thickness of the LaAlO$_3$ overlayer. However, unlike the non-strained state, the conductivity versus thickness of LaAlO$_3$ had a gradual rather than an abrupt change at the critical thickness. For instance, in case of LaAlO$_3$/SrTiO$_3$/LSAT measurable conductivity was detected at 10 unit cells LaAlO$_3$ thickness, but it did not saturate until 20 unit cells. There is however a clear trend of increasing LaAlO$_3$ critical thickness with increasing compressive biaxial in-plane strain.

Figure 2B shows the carrier concentration at each strain state above the critical thickness of LaAlO$_3$. Similar to the critical thickness of LaAlO$_3$ layer, we find nearly the same carrier concentration at both near-zero strain states, LaAlO$_3$ on SrTiO$_3$ bulk single crystal and LaAlO$_3$ on quasi-single crystal (001) SrTiO$_3$ template on Silicon. The saturation carrier concentration (above the critical thickness) decreased with increasing compressive strain Although LaAlO$_3$/SrTiO$_3$ interfaces on DSO and GSO were grown and treated in the same manner, the interfaces were not conducting within our measurement limit at any thickness of LaAlO$_3$ overlayer in these tensile-strained films.

Our experimental results indicate that tensile-strained SrTiO$_3$ destroys the conducting interfacial 2DEG, while compressive-strained SrTiO$_3$ preserves the 2DEG, but with decreased interfacial carrier concentration. The maximum carrier concentration at the SrTiO$_3$ unstrained state suggests that it is the strain-dependence of SrTiO$_3$ properties that control the 2DEG. It has been predicted theoretically that free-standing biaxially strained SrTiO$_3$ under electrical short-circuit boundary conditions can develop an electric polarization (20,21). Compressive strain is predicted to produce an [001] (out-of-plane) polarization, and tensile strain to produce a [110] (in-plane) polarization. Experimental evidence suggests a more complex picture, with many strain-states resulting in a relaxor behavior at room temperature (11,,22) without a stable switchable polarization.

However we expect the strain-induced STO properties to be altered by the LaAlO$_3$ overlayer. Observations from TEM (23,24), synchrotron radiation x-ray scattering (25,26), and tunneling (27), indicate that in strain-free SrTiO$_3$ a few unit cells near the LaAlO$_3$ interface have ferroelectric-like structural distortions with local polarization pointing away from the interface, and decreasing in magnitude with distance from the interface (28). Biaxial compressive strain induces a tetragonal distortion along 001, which would enhance this polarization, potentially uniformly polarizing the SrTiO$_3$ throughout its thickness. (29).

Figure 3 schematically compares the strained and unstrained systems. In the unstrained system positively charged (LaO)$^+$ atomic layers and negatively charged (AlO$_2$)$^-$ atomic layers create an average polarization whose positive bound charge resides at the interface, as shown schematically in the left panel of Fig. 3A. This polarization charge is responsible for the intrinsic electric field $E_0$ in LaAlO$_3$ (shown by arrow in Fig. 3A) resulting in an electric potential difference between the LaAlO$_3$ surface and the LaAlO$_3$/SrTiO$_3$ interface that increases with LaAlO$_3$ layer thickness. Above the LaAlO$_3$ critical thickness, charge is transferred to the LaAlO$_3$/SrTiO$_3$ interface (shown by a blue filling) to avoid this polarization catastrophe.

The compressively strained SrTiO$_3$ layer contains polar displacements of the Ti$^{4+}$ ions with respect to the O$^{2-}$ ions, shown in Fig. 3B for the case of uniform polarization. These displacements are responsible for a polarization $P$ pointed away from the interface (indicated by an arrow at the bottom of the left panel of Fig. 3B). The polarization orientation is determined by the presence of the LAO layer and is likely not switchable. The polarization produces a negative bound charge at the LaAlO$_3$/SrTiO$_3$ interface (indicated in the left panel of Fig.3B) that creates an additional electric field in LaAlO$_3$ equal to $P/\varepsilon$, where $\varepsilon$ is the dielectric constant of LaAlO$_3$, that opposes the intrinsic electric field $E_0$. The presence of polarization in the compressively strained SrTiO$_3$ layer reduces the total electric field in LaAlO$_3$ and hence enhances the critical thickness necessary to create a 2DEG at the LaAlO$_3$/SrTiO$_3$ interface due to the polarization catastrophe effect. Above this critical thickness, the mobile interfacial carrier concentration would be reduced by the interfacial bound charge (12,13).

In order to quantify these effects we have completed first-principles calculations of the LaAlO$_3$/SrTiO$_3$ bilayer under various strain states based on density functional theory (DFT), as described in *Materials and Methods*. Fig. 4 shows calculated ionic displacements for the unstrained and 1.2% compressively strained (LaAlO$_3$)$_3$/(SrTiO$_3$)$_5$ structures. It is seen that in the unstrained case polar Ti-O displacements in the SrTiO$_3$ layer are very small, consistent with the previous calculations (33). The in-plane 1.2% compressive strain produces sizable ionic displacements, polarizing the SrTiO$_3$ layer. The calculation predicts that the induced polarization is oriented away from the interface and is not switchable. The magnitude of the polarization is $P \approx 0.18$ C/m$^2$, as found from the known polar displacements in the strained SrTiO$_3$ layer using the Berry phase method (30,31).

The critical thickness $t_c$ in the presence of a STO polarization can be estimated as follows:

$$t_c = \delta\mathcal{E}/eE \qquad (1)$$

where $\delta\mathcal{E} = e\mathcal{E}_g + (\mathcal{E}_{VBM}^{STO} - \mathcal{E}_{VBM}^{LAO})$, $\mathcal{E}_g$ is the band gap of SrTiO$_3$, $\mathcal{E}_{VBM}^{STO}$ and $\mathcal{E}_{VBM}^{LAO}$ are the valence band maxima (VBM) of SrTiO$_3$ and LaAlO$_3$ respectively, and $E$ is the electric field in LaAlO$_3$. The latter is reduced from the intrinsic value of $E_0$ due to polarization $P$ of SrTiO$_3$ so that

$$E = E_o - \frac{P}{\varepsilon_{LAO}}, \qquad (2)$$

where $\varepsilon_{LAO}$ is the dielectric constant of LaAlO$_3$. Due to the reduced electric field in LaAlO$_3$ in the presence of the SrTiO$_3$ polarization, the critical thickness (1) is enhanced. The intrinsic electric field $E_0$ can be estimated from the experimentally measured critical thickness $t_c^0 = 4$ u.c. for the unstrained system. Taking into account the experimental band gap of STO $\mathcal{E}_g = 3.2$ eV and the CBM offset between SrTiO$_3$ and LaAlO$_3$ $\mathcal{E}_{VBM}^{STO} - \mathcal{E}_{VBM}^{LAO} = 0.35$ eV, (32) we find that $\delta\mathcal{E} = 3.55$ eV. Using the relationship

$$\delta\mathcal{E} = eE_o t_c^0 \qquad (3)$$

we obtain that $E_0 \approx 0.23$ V/Å which is consistent with our first-principles calculation predicting $E_0 \approx 0.22$ V/Å and calculations by others (33,34,35). Using Eqs. (1-3) we obtain

$$t_c = \frac{t_c^0}{1 - \dfrac{P}{\varepsilon_{LAO} E_0}} . \qquad (4)$$

Using the calculated polarization value $P \approx 0.18$ C/m$^2$ for 1.2% compressive strain in the STO layer, and the calculated electric fields in the LaAlO$_3$ and SrTiO$_3$ layers in the strained LaAlO$_3$/SrTiO$_3$ system, we estimate the dielectric constant of the LaAlO$_3$ grown on 1.2% compressively strained SrTiO$_3$ to be $\varepsilon_{LAO} \approx 18\varepsilon_o$. This value is consistent with that obtained from the induced polarization of 0.34 C/m$^2$ in the LaAlO$_3$ layer, as is estimated from the calculated ionic displacements using the Berry phase method. (We note that the estimated value of the dielectric constant of the unstrained LaAlO$_3$ is $\varepsilon_{LAO} \approx 24\varepsilon_o$ which is consistent with the previously found result (36)). Using. Eq. (4) and the dielectric constant $\varepsilon_{LAO} \approx 18\varepsilon_o$ we obtain $t_c \approx 9$ u.c. This value is higher than the critical thickness (4 u.c.) for the unstrained system, and is consistent with the experimental result for the 1.2% strained LaAlO$_3$/SrTiO$_3$ structure. It is relevant to the experimental situation where it is expected that the surface polarization charge in LaAlO$_3$ is screened by adsorbents and the bottom polarization charge in the strained SrTiO$_3$ is screened by defects. In the structural model used in our DFT calculation the SrTiO$_3$ polarization is screened by charge transferred to the SrTiO$_3$ surface.

For the case of tensile strain in the STO layer, our experiments indicate that there is no conducting 2DEG for biaxial tensile strains above 1.1%. Free-standing $SrTiO_3$ has been predicted at zero temperature to develop an in-plane polarization in the (110) direction under biaxial tensile strain. Experiment suggests that, at room temperature, relaxor behavior, with nanoscale polar regions that can be aligned in an electric field, occurs in many tensile strained $SrTiO_3$ samples. Stabilization of a uniform in-plane polarization by the $LaAlO_3$ layer does not seem likely. If such nanoscale regions near to the interface were present in our samples, bound charge at polarization discontinuities between random nanopolar regions would tend to be locally screened by carriers at the 2DEG interface. This would lead to localization of these carriers, preventing us from observing conduction in these samples.

Another aspect is strain in the $LaAlO_3$ overlayer. The bulk pseudocubic lattice constant of $LaAlO_3$ is 3.791 Å, so that coherent $LaAlO_3$ even on unstrained $SrTiO_3$ has a 3% tensile strain. Growing the bilayer on a GSO substrate results in 4.5% tensile strain in the $LaAlO_3$ layer. An NGO substrate reduces the $LaAlO_3$ strain to 1.8% tensile, but for all substrates used the LAO layer is under tensile strain. Our transmission electron microscopy analysis of these samples indicates that the $LaAlO_3$ layer on $SrTiO_3$ is fully coherent when grown on LSAT (2 % $LaAlO_3$ tensile strain) and STO, but that growth on DSO (leading to 4 % $LaAlO_3$ tensile strain) results in partial relaxation of the $LaAlO_3$. Such defect incorporation might alter the conduction properties of the interface. However, the $SrTiO_3$ layer on Si (grown by MBE) is almost fully relaxed, and the bilayer shows a fully conducting interfacial 2DEG, but with lower mobility. This suggests that such defects do not destroy the 2DEG. Large tensile strain in $LaAlO_3$ has been predicted (37) to alter the Al-O bond lengths, which could affect the electronic structure.

We have demonstrated that properties of the 2DEG formed at the $LaAlO_3/SrTiO_3$ interface can be controlled by epitaxial strain. Both the critical thickness of the $LaAlO_3$ overlayer required to generate the 2DEG and the carrier concentration of the 2DEG depend on the strain of the $SrTiO_3$ layer. Compressive strain increases the critical thickness and decreases the saturated carrier concentration. Our DFT calculations indicate that a strain-induced polarization stabilized by the $LaAlO_3$ overlayer is responsible for these changes. Changes in critical thickness and carrier concentration estimated from the DFT calculations are in agreement with the experimental data.

The dependence of 2DEG properties at the $LaAlO_3/SrTiO_3$ interface on the strain state opens a new correlation between strain-induced polarization and the electrical properties of oxide interfaces. We believe that such strain engineering can be very useful for oxide 2DEG device applications, and the relation between strain and 2DEG properties provides a new tool in the manipulation of oxide interfacial 2DEGs.

**Materials and Methods**

Epitaxial LaAlO$_3$ and STO thin films were grown on (001) LSAT, (110) NdGaO$_3$, (110) GdScO$_3$, and (110) DyScO$_3$ substrates by PLD. To grow heterostructures by PLD, substrates were attached to a resistive heater and positioned 5.0~6.0 cm from the target. A KrF excimer laser (248 nm) beam was focused on a stoichiometric LaAlO$_3$ and SrTiO$_3$ single crystal target to an energy density of 2.0~2.5 J/cm$^2$ and pulsed at 3~5 Hz. SrTiO$_3$ templates were grown at substrate temperatures ranging from 650 to 850 °C and oxygen pressures of 10-100 mTorr. Before deposition, low miscut (<0.05°) LSAT, NGO, DSO, GSO substrates were treated by a modified buffered HF etch and annealed in oxygen at 1000~1100°C for 2~12 hours to create atomically smooth surfaces with unit cell stepThe PLD system is equipped with high-pressure RHEED, which enabled atomic layer controlled growth and *in situ* monitoring during the growth. SrTiO$_3$ templates were etched using buffered HF acid for 30~90 seconds to maintain Ti-termination after growth SrTiO$_3$ layer.. LaAlO$_3$ films were grown at 550 °C at oxygen pressures of 10$^{-3}$ mbar and cooled down to room temperature at the same oxygen pressure.

The three-dimensional strain state of the films was determined using high-resolution four-circle x-ray diffraction (Bruker D8 advance). The microstructure and interfacial structure of the samples were characterized by cross-sectional transmission electron microscopy (TEM). Surface of films were imaged by AFM (Veeco).

After the growth, Al contacts were made by wire bonding near the four corners of the sample for van der Pauw electrical characterization. A Keithley 2700 sourcemeter combined with a 2400 switch matrix multimeter was used for the van der Pauw measurements of conductance and carrier concentration. The sheet resistance was calculated by fitting slopes to the four point IV curves measured between the four combinations of contacts. The nominal sheet carrier concentration was determined from the Hall coefficient as $n_{2D}=-t/R_H e$ where $t$ is the film thickness, $R_H$ is the Hall coefficient, and e is the charge of an electron. The mobility was determined from the sheet resistance $R_\square$ and sheet carrier concentration $n_{2D}$ as $\mu=1/en_{2D}R_\square$.

Density functional calculations were performed within the local density approximation (LDA) using the VASP method (38,39), similar to the calculations performed previously (40). We considered a LaO/TiO$_2$-interfaced (LaAlO$_3$)$_n$/(SrTiO$_3$)$_m$ bilayer (where *n* and *m* are the numbers of unit cells of LaAlO$_3$ and SrTiO$_3$ respectively), as a model system. The LaAlO$_3$/SrTiO$_3$ bilayer was placed in a LaAlO$_3$/SrTiO$_3$/vacuum/SrTiO$_3$/LaAlO$_3$/vacuum supercell, where the doubled bilayer was used to avoid an unphysical electric field in vacuum which otherwise would occur due to the potential step within the LaAlO$_3$ layer and periodic boundary conditions of the supercell calculations. The in-plane lattice constant of the unstrained superlattice was fixed to the calculated bulk lattice constant of SrTiO$_3$, *i.e. a* = 3.871 Å. For the strained systems the in-plain lattice constant was constrained to be by a certain percentage smaller than the bulk one. To reduce the effect of the SrTiO$_3$ surface on atomic structure and ionic displacements within the SrTiO$_3$ layer we used a boundary condition according to which the atomic positions within one unit cell on the SrTiO$_3$ surface were fixed to be the same as in the respectively strained bulk SrTiO$_3$. The latter were computed separately for the unstrained and strained bulk SrTiO$_3$. All the other atoms in the superlattices were relaxed.

**Acknowledgements**

This work was supported by the National Science Foundation under Grant No. DMR-0906443, and a David and Lucile Packard Fellowship (C.B.E.). The research at University of Nebraska was supported by the Materials Research Science and Engineering Center (NSF Grant No. DMR-0820521) and the Nebraska Research Initiative. The work at the University of Michigan was supported by DMR-0907191, DoE/BES DE-FG02-07ER46416, and NSF/DMR-0723032.

**Figure Legends**

Figure 1. Structural characterization of hetero structures. (A) Schematic diagram of grown structures. Thickness of LaAlO$_3$ layer was varied from 1 to 30 unit cells on STO on LSAT, NGO, Si, DSO, GSO substrate, (B) RHEED intensity oscillations for the growth of LAO and STO on LSAT substrate. The insets show the RHEED pattern at the end of the LAO and STO growth. (C) High-resolution transmission electron microscopy image of LAO on STO on LSAT. Clear intensity differences between separate LaAlO$_3$ and SrTiO$_3$ layers in both cases and images showing distinctly the individual La and Sr atoms. (D) AFM images of LAO on STO on DyScO$_3$, substrates.

Figure 2. Effect of strain on 2DEG. (A) Critical thickness of LaAlO$_3$ under bi-axial strain While others samples had 50 unit cell-thick SrTiO$_3$ layer, LAO/STO/Si had 100nm-thick STO to get nominally unstrained STO layer on silicon. Conductivity verses thickness of LaAlO$_3$ in LAO/STO interface on various substrates was represented in inset. (B) Sheet resistance at LAO/STO interface under bi-axial strain. Sheet resistances in tensile strain state were above our measurement limit

Figure 3. Calculated atomic structure of unstrained (A) and compressively strained (B) LaAlO$_3$(3u.c.)/SrTiO$_3$ system. In figure B Ti-O and Sr-O displacements are amplified by a factor of eight as compared to the calculated results for visual comprehension. The left and right panels show schematically the 2DEG formation and the effect of the polarization P in the strained SrTiO$_3$ on the 2DEG as described in text.

Figure 4. B (Ti, Al) cite atom – oxygen (O) atom displacements in the unstrained (squares) and 1.2% compressively strained (circles) (LaAlO$_3$)$_3$/(SrTiO$_3$)$_5$ structure.

**Table Legend**

Table 1. Results from high-resolution x-ray diffraction measurements on the films at room temperature are given. The in-plane (a) and out-of-plane (c) lattice

constants and lattice mismatch between the SrTiO$_3$ films and single crystal substrates on average of two orthogonal directions. The a- and c-lattice parameters of single-crystalline SrTiO$_3$ are 3.905Å. All SrTiO$_3$ templates were fully coherent except STO/Si (12). (002), (101) of SrTiO$_3$ and cubic substrates, LSAT, Silicon (200)$_{pseudo-cubic}$ of (101)$_{pseudo-cubic}$ of orthorombic substrate, GdScO$_3$ and DyScO$_3$, NdGdO$_3$ were observed to determined in-plane and out-of-plane lattice parameters.

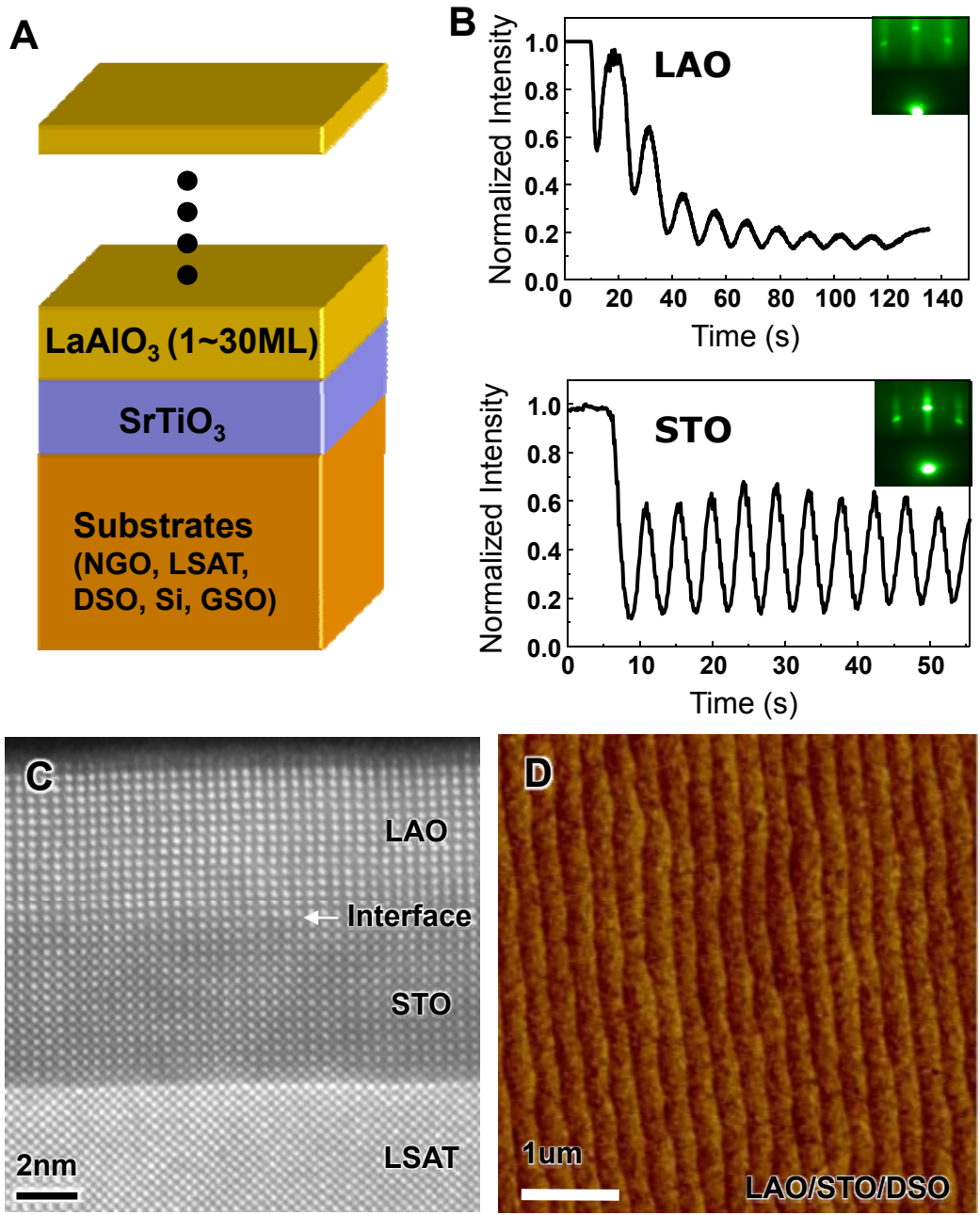

Figure 1, C.W. Bark et al,

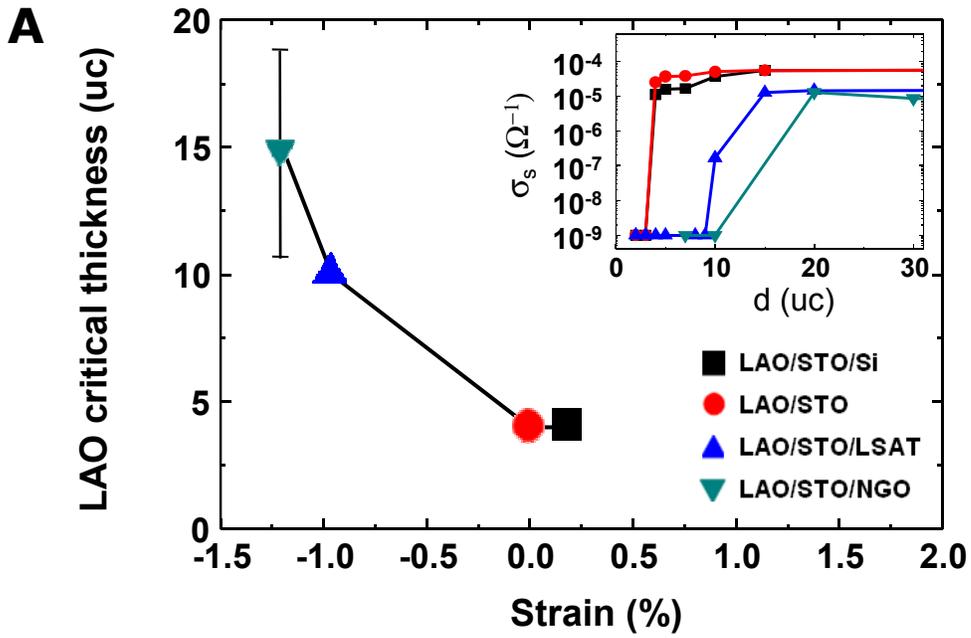

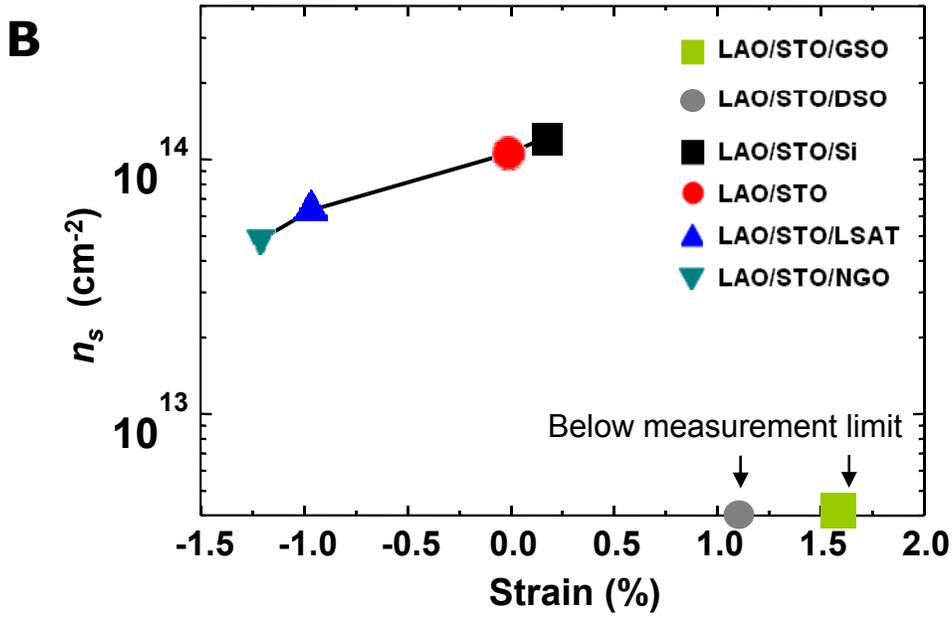

Figure 2, C.W. Bark et al

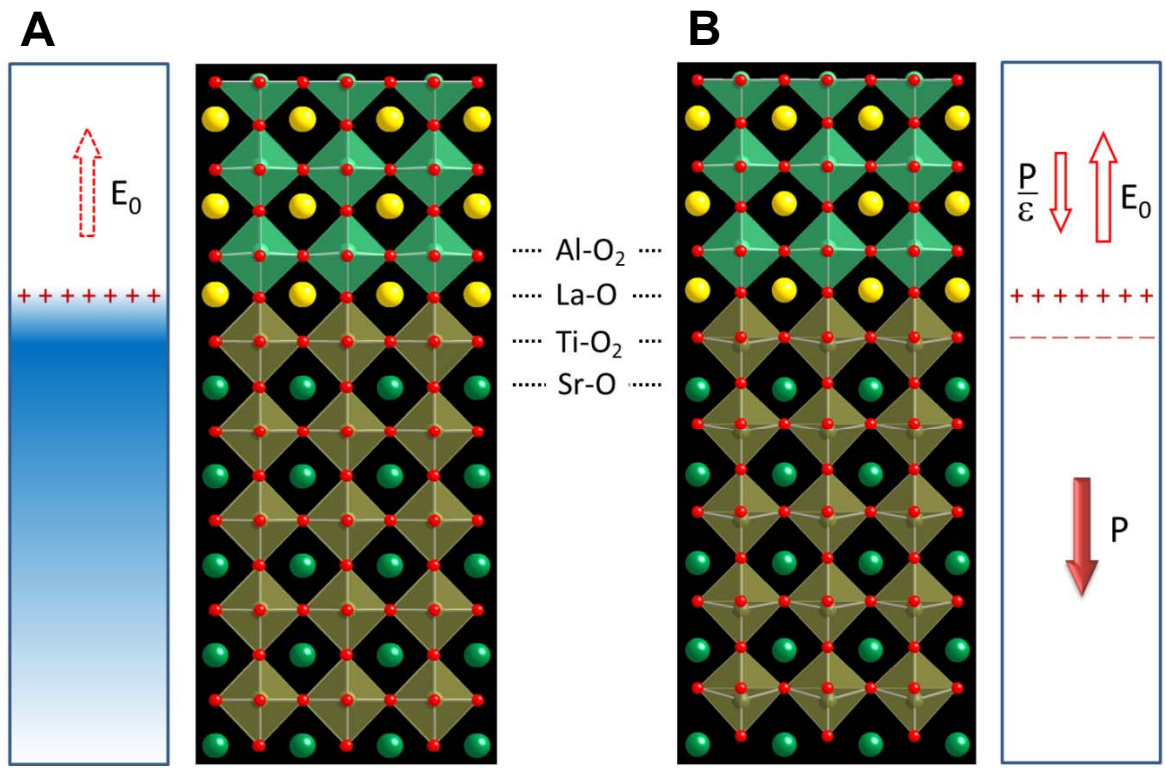

Figure 3, C.W. Bark et al

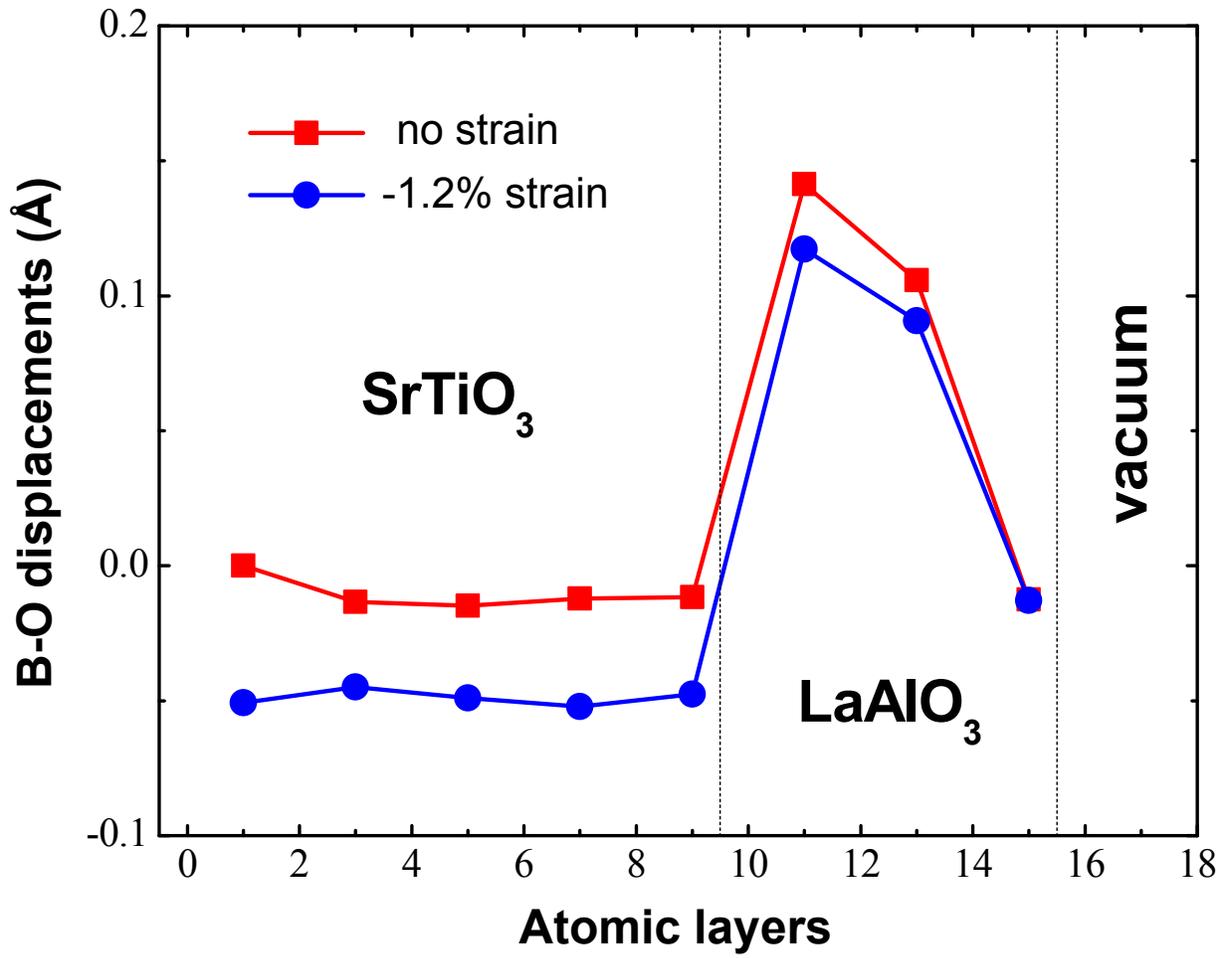

Figure 4, C.W. Bark et al

|  | a (Å) | c (Å) | bi-axial lattice mismatch |
|---|---|---|---|
| *LaAlO3 (10uc) on SrTiO3 (50uc) on NdGaO3* | | | |
| SrTiO3 | 3.860 | 3.964 | -1.21% |
| NGO | 3.859 | 3.866 | |
| *LaAlO3 (10uc) on SrTiO3 (50uc) on LSAT* | | | |
| SrTiO3 | 3.868 | 3.940 | -0.96% |
| LSAT | 3.869 | 3.387 | |
| *LaAlO3 (10uc) on SrTiO3 (120nm) on Si* | | | |
| SrTiO3 | 3.911 | 3.985 | 0.15% |
| Si | 3.840 | 3.840 | |
| *LaAlO3 (10uc) on SrTiO3 (20uc) on DyScO3* | | | |
| SrTiO3 | 3.944 | 3.939 | 1.11% |
| DyScO3 | 3.944 | 3.939 | |
| *LaAlO3 (10uc) on SrTiO3 (20uc) on GdScO3* | | | |
| SrTiO3 | 3.964 | 3.875 | 1.59% |
| GdScO3 | 3.963 | 3.967 | |

Table 1. Bi- axial strain of SrTiO3 templates.